\documentclass[aps,prl,twocolumn]{revtex4-1}
\usepackage[hidelinks]{hyperref}
\usepackage{mathtools}
\usepackage{braket}
\usepackage{amsfonts}
\usepackage{amsmath}

\def\be{\begin{equation}}
\def\ee{\end{equation}}
\def\bea{\begin{eqnarray}}
\def\eea{\end{eqnarray}}

\def\be{\begin{equation}}
\def\ee{\end{equation}}
\def\bea{\begin{eqnarray}}
\def\eea{\end{eqnarray}}

\def\ms{\mathsf}

\def\be{\begin{equation}}
\def\ee{\end{equation}}
\def\bea{\begin{eqnarray}}
\def\eea{\end{eqnarray}}

\def\be{\begin{equation}}
\def\ee{\end{equation}}
\def\bea{\begin{eqnarray}}
\def\eea{\end{eqnarray}}

\newcommand{\td}{\ensuremath{{\text{d}}}}

\begin{document}

\title{Black Hole Entropy in General Relativity}

\author{Thomas C. Bachlechner}
\affiliation{Department of Physics, University of California San Diego, La Jolla, CA 92093}
\vskip 4pt

\begin{abstract}
The Bekenstein--Hawking formula relates the  black hole entropy and horizon area. Semiclassical entropy computations have relied on an action principle that fixes a gauge dependent and classically unobservable boundary three-geometry and renders elusive a precise physical notion of both energy and entropy in de Sitter backgrounds. Instead, we impose gauge invariant boundary conditions and report the background independent action for black hole formation. Assuming standard arguments for the relation between the action and entropy, we reproduce the Bekenstein--Hawking formula and motivate a quantization of the phase-space volume. This background independent approach applies to spacetimes of arbitrary energy density and enables a radically conservative framework for semiclassical gravity.

\end{abstract}

\maketitle

\section{Introduction}
Consider a spherically symmetric black hole of invariant mass $\ms M$ in four dimensional Einstein gravity \cite{Einstein:1916vd}. Physical black holes are formed by the collapse of matter. An observer of the black hole in asymptotically flat space notes a horizon at $\ms R_\text{h}=2 G \ms M$, but has no access to the initial configuration. It has been argued that this ignorance is quantified by the black hole entropy ${\cal S}$ \cite{Bekenstein:1972tm,Bekenstein:1973ur,Bardeen:1973gs,Hawking:1974sw,Zurek:1985gd}. 

Semiclassical computations of the entropy require an action principle. Gibbons, Hawking and York proposed a commonly used action which assumes a fixed boundary three-geometry, and yields a well posed variational principle \cite{Arnowitt:1962hi,York:1972sj,Regge:1974zd,Gibbons:1976ue,Brown:1992br,Brown:1992bq}. However, these boundary conditions are not invariant under coordinate reparametrization and require an additional, classically unobservable prescription to define an energy and time-translation structure. Both the energy and black hole entropy would be ill-defined in a de Sitter background, which  constitutes a severe but largely ignored issue of principle.

\begin{figure}
\centering
\includegraphics[width=0.35\textwidth]{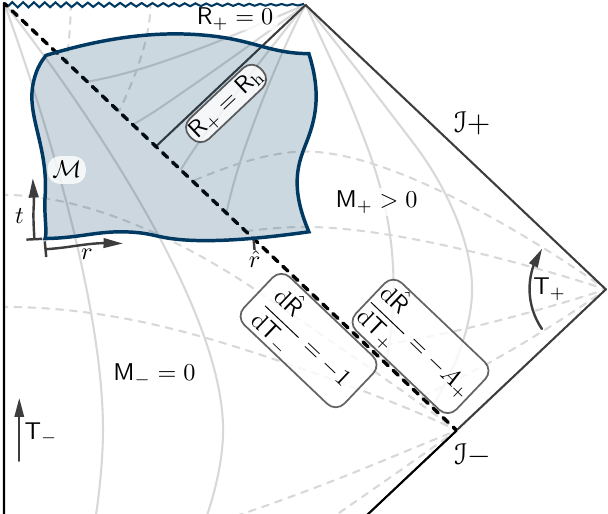}
\caption{\small Conformal diagram of black hole formation via the collapse of a thin, null shell. Lines of constant Killing time and radius are shown as dashed and solid gray lines, respectively.\label{figurer}}
\end{figure}

In this work we instead restrict to reparametrization invariant boundary conditions that yield a prescription and coordinate independent action principle for general relativity  \cite{tbgravity}. This conceptually different framework has an unambiguous time-translation structure that resolves the issues that arise in closed spacetimes. Let us briefly sketch the intuitive computation that we perform in detail below. Up to  an irrelevant real part, the gravitational action for black hole formation from the collapse of a thin shell reads \cite{tbvacuumdecay}
\be\label{domwall1}
S=-\int \td \ms M \td\ms T+\int\hat{P} \td \hat{\ms R}\approx -2\times\int  {\ms R\over \ms R-\ms R_\text{h}} {\td \ms R \td {\ms M}}\,.
\ee
We illustrate the process in Figure \ref{figurer}. The action (\ref{domwall1}) is conceptually new, background independent and easy to understand. The first term implies that the invariant mass $\ms M$ is the Hamiltonian associated with Schwarzschild  time $\ms T$ translations within the static spacetime away from the shell. The second term implies that $\hat{P}$ is the momentum conjugate to the shell radius $\hat{\ms R}$. Near the (null) horizon we have $\td \ms R/\td \ms T\approx 1-\ms R_\text{h}/\ms R$, and Hamilton's equation for a collapsing shell near the horizon reads $\td {\ms  M}/\td \hat{P}=\td \hat{\ms R}/\td {\hat{\ms T}}\approx\ms R_\text{h}/ \hat{\ms R}-1$, adding up to  the right hand side of (\ref{domwall1}). Integrating over the pole at the horizon gives the action
\be
{2}\times  \text{Im S}={\Delta \mathbb A_\text{h}\over 4 G} \,,
\ee
which depends only on the change of the black hole horizon area $\mathbb A_\text{h}=4\pi \ms R_\text{h}^2$. This computation is independent of the asymptotic spacetime and applies for an arbitrary cosmological constant. We discuss the thermodynamic properties of spacetimes with positive vacuum energy density separately in \cite{dsthermo}.

It has been argued \cite{KeskiVakkuri:1996xp} (see also \cite{Parikh:1999mf}) that the probability for emission of Hawking radiation at an energy that causes a change in the horizon area of $\Delta \mathbb A_\text{h}$ is
\be\label{tunnelp}
{\cal P}=e^{{2}\times  \text{Im S}/\hbar}\,,
\ee
which results in an approximately thermal spectrum of radiation as if the black hole had an entropy $\cal S$ and phase-space volume $\Omega=e^{{\cal S}/k}$ given by the Bekenstein--Hawking formula \cite{Bekenstein:1973ur,Hawking:1974sw}
\be\label{bhformula}
 {\cal S}=-k\times \log ({\cal P})={k\over \hbar}{ \mathbb A_\text{h}\over 4 G}\,.
\ee
Switching to the canonical ensemble, the  probabilities (\ref{tunnelp}) imply that a black hole of temperature $\beta^{-1}=\hbar/4\pi k \ms \langle R_\text{h}\rangle$ radiates according to a Planck spectrum \cite{Bardeen:1973gs,Gibbons:1976ue, Parikh:1999mf}.

We can heuristically relate the quantum mechanical (momentum) uncertainty relation to the quantization of the phase-space volume $\Omega$. The longest semiclassical timescale available to a system with the black hole entropy is $\Delta T\approx 2\pi \ms R_\text{h} e^{\mathbb A_\text{h}/4 G}$, which coincides both with the time for   quantum emission of a single state of mass $\ms M$, and the recurrence time of a thermal system. The quantum  uncertainty relation, $\Delta \ms M \Delta T \gtrsim \hbar/2$, implies that we cannot possibly prepare a state of mass less than $\Delta \ms M$ to add to the black hole. This would demand the quantization of the phase-space volume $\Omega = e^{{\cal S}/k}$,
\be
\Delta\Omega\ge 1\,.
\ee
This relation is particular to black holes \footnote{We are not aware of a reference for this observation, but given its simplicity it likely has previously appeared.}.

We see that the coordinate invariant framework for semiclassical general relativity enables  a prescription independent  evaluation of the  black hole entropy. If realized in nature, it constitutes a qualitatively new approach in which coordinates are mere redundancies, three-geometries remain unfixed, and some  puzzles about gravity deserve reconsideration.

\section{Setup}
We now return to natural units $k=\hbar=1$. For simplicity we are interested in the semiclassical description of a thin and spherical domain wall of tension $\sigma$, evolving in a general (spherically symmetric) spacetime, potentially containing a black hole \cite{fgg,fmp1,fmp2,Kraus:1994by, KeskiVakkuri:1996xp,Parikh:1999mf,runaways}. Our results will be independent of the details of the in-falling matter, or the cosmological constant. There always exists a gauge in which the metric takes the form
\be
	\label{staticmetric}
\td s_\pm ^2 = -A_\pm \td {\ms  T}^2 + A^{-1}_\pm  \td {\ms  R}^2 + {\ms  R}^2 \td\Omega_2^2 \,,
\ee
where ${\ms  T}$ denotes the Schwarzschild (or Killing) time of the static spacetimes, the signs $-/+$ denote the shell interior/exterior, and we defined
\be
A_\pm=1-{2 G \ms M_\pm\over \ms R}\,.
\ee
The invariant masses observed on either side of the shell and the horizon radii are denoted by $\ms M_\pm$ and $\ms R_{\text{h}\pm}$, respectively. While (\ref{staticmetric}) is a familiar gauge choice, general relativity requires coordinate invariance. In order to accommodate arbitrary diffeomorphisms of the $r$ and $t$ coordinates, we work with the general spherically symmetric metric, which can  be written as \cite{tbgravity}
\be\label{metric}
ds^2=\left(A^{-1}{\ms R'^2}-A{\ms T'^2}\right)\left[(\td r+ \ms N_r\td t)^2- \ms N_t^2 \td t^2\right]+ \ms R^2 \td \Omega_2^2,
\ee
where  we promoted the variables $\ms T$, $\ms R$ and $\ms M$ appearing in (\ref{staticmetric}) to coordinate dependent functions, such that the lapse $\ms N_{t}$ and the shift $\ms N_{r}$ can absorb arbitrary reparametrizations. Primes and dots denote derivatives in the coordinates $r$ and $t$, respectively. The general metric (\ref{metric}) becomes the Schwarzschild metric (\ref{staticmetric}) in the gauge $\ms T=t$ and $\ms R=r$, where $\ms N_r=0$ and $\ms N_t=A$. We use the radial coordinate $r$ to distinguish the interior ($r<\hat{r}$) from the exterior ($r>\hat{r}$) of the shell located at $\hat{r}$.

The equations of motion for the shell and spacetime follow from the Euler--Lagrange equations of the action
\be\label{gravityaction}
S=\underbrace{ \int_{S_2}\td \Omega_2 \int_{\cal M}\td t\td r\, \sqrt{g} {{\cal R}\over 16\pi G}}_{S_\text{G}+\dots} -\underbrace{\sigma\int_\text{wall} \td^3{\cal A}}_{-S_\text{Shell}}+S_\text{BT},
\ee
where ${\cal R}$ is the Ricci scalar, ${\cal M}$ denotes the relevant coordinate region, $\sigma$ is the shell tension and  ${\cal A}$ the world-volume. $S_\text{BT}$ and ellipses denote  total derivative actions which ensure that the variational principle $\delta S=0$ implies the equations of motion. Without loss of generality we choose the manifold $\cal M$ as bounded by constant coordinates,
\be
{\cal M}=\{\,(t,r)~:~t_{\text{i}}\le t\le t_{\text{f}}\,,~
r_{\text{min}}\le r\le r_{\text{max}}\,\}\,.
\ee
 
We can easily impose boundary conditions that transform trivially under diffeomorphisms by fixing the invariant mass, the two-sphere radius and the shell-location at the boundary
\be\label{bcs}
\delta{\hat{r}}|_{\partial \cal M}=\delta{\ms M}(t,r)|_{\partial \cal M}=\delta \ms R(t,r)|_{\partial \cal M}=0\,. 
\ee
These boundary conditions differ from previous approaches to quantum gravity that fix the classically unobservable and coordinate dependent boundary three-geometry  \cite{Arnowitt:1962hi,York:1972sj,Regge:1974zd,Gibbons:1976ue,Brown:1992br,Brown:1992bq}. We will further comment on the differences in the discussion below.

\section{Action for Black Hole Formation}
Let us now evaluate the action (\ref{gravityaction}) for a collapsing shell to change the black hole horizon area by an amount $\Delta \mathbb A_\text{h}$, as shown in Figure \ref{figurer} for a null shell with vacuum interior. 

As a first step we discuss the gravitational action and ignore the shell, so let us drop the subscripts $\pm$ for now. Demanding that the boundary terms appearing in the variation of the action $S_\text{G}$ vanish under Dirichlet boundary conditions for the invariant mass $\ms M$ and two-sphere radius $\ms R$ yields the canonical action for isotropic general relativity \cite{tbgravity}
\be\label{gravaction}
S_\text{G}=\int_{\cal M}\td t  \td r ~ \pi_{\ms M} \dot{\ms M} + \pi_{\ms R} \dot{\ms R} - {\ms N_t} {\cal H}^\text{G}_t - {\ms N_r} {\cal H}^\text{G}_r \,,
\ee
where $\pi_{\ms M}= -\ms T'$  is the momentum density conjugate to $\ms M$ and we defined the Hamiltonian densities
\be
{\cal H}^\text{G}_r=\pi_{\ms R} \ms R'+\pi_{\ms M} \ms M'\,,~~{\cal H}^\text{G}_t=A \pi_{\ms M}\pi_{\ms R}+A^{-1}{\ms M'\ms R'}\,.
\ee
While we do not allow for diffeomorphisms that mix the angular with the $t$ or $r$ coordinates, this mini-superspace assumption yields an analytically tractable and commonly used toy model for isotropic gravity \cite{Almheiri:2014cka}. More precisely, in this work we study the 1+1 dimensional dilaton gravity theory that arises from the dimensional reduction of four dimensional Einstein gravity on a two-sphere. It is easy to check that the boundary variations $\delta S_\text{G}|_{\partial \cal M}$ vanish, while no boundary conditions for the gauge dependent Lagrange multipliers $\ms N_{t,r}$ are required or allowed. This observation confirms that the conserved charges (like the total Hamiltonian) associated with  diffeomorphisms vanish, as required in general relativity. Variations of the action with respect to the momentum densities give the relations between momenta and velocities $\dot{\ms R}\,, \dot{\ms M}$, while variations with respect to the lapse and the shift impose the Hamiltonian constraints ${\cal H}^{\text{G}}_{t,r}=0$. These constraints can be written as $\pi_{\ms R}=\ms M'=0$, and imply the classical action
\be\label{classicalg}
S_\text{G}=- \int_{\cal M} \delta{\ms M} \td r ~ \ms T'\,.
\ee
The  action (\ref{classicalg}) unambiguously demonstrates that the invariant mass $\ms M$ is the Hamiltonian associated to the Schwarzschild (Killing) time $\ms T$ translations. These translations  are coordinate independent, so their associated Noether charge can be non-zero consistent with diffeomorphism invariance. The Killing time gradient $\ms T'$ has a pole at the horizon. To see this, we pick a convenient, non-singular gauge where $g_{rr}=1$ and find from the metric (\ref{metric})
\be\label{momM}
\pi_{\ms M}=-\ms T'=\eta_\pi A^{-1}{\sqrt{\ms R'^2-A }}\,,
\ee
where $\eta_\pi=\pm 1$ parametrizes the arbitrary sign choice for the Killing time $\ms T$. Choosing coordinates $\ms R=r$ in the vicinity of the horizon $\ms R\approx\ms R_\text{h}$, where $A\approx 0$ we find the imaginary part of the gravitational action
\be\label{gravim}
\text{Im}~S_\text{G}\approx\text{Im}~\int_{\ms M(t_i)}^{\ms M(t_f)}\td\ms M\int_{r_\text{min}}^{r_\text{max}}\td r { \eta_\pi r \over r-{2 G \ms M} }= -\eta_\pi{\Delta \mathbb A_\text{h}\over 16G}\,,
\ee
where we deformed the integration contour by a small amount into the complex plane as to ensure convergence of the path integral and defined the  change in the black hole horizon area  $\Delta \mathbb A_\text{h}\equiv 16 \pi G^2[\ms M^2(t_f)-\ms M^2(t_i)] $.

With (\ref{gravim}) we found the purely gravitational contribution to the action for the black hole formation process. We now turn to evaluating the contribution of the shell, $S_\text{Shell}+S_\text{BT}$.  The action of a thin shell with surface tension $\sigma$ is given with the metric (\ref{metric}) as 
\bea\label{ssshell}
S_\text{Shell}&=&-\int_{t_\text{i}}^{t_\text{f}} \td t\, \hat{m}\sqrt{g_{rr}(\ms N_t^2-[\dot{\hat{r}}+\ms N_r]^2) }
\\
 &=&\int_{t_\text{i}}^{t_\text{f}} \hat{p}\dot{\hat{r}}-\int_{r_\text{min}}^{r_\text{max}} dr~ ({\ms N_t} {\cal H}_{t}^\text{Shell} +{\ms N_r} {\cal H}_{r}^\text{Shell})~ \td t\,,\nonumber
\eea
where we defined the domain wall radius and rest mass as $\hat{\ms R}\equiv R(t,\hat{r})$ and $\hat{m}\equiv 4\pi\sigma \hat{\ms R}^2$, respectively. We recover a null shell in the limit $\hat{m}\rightarrow 0$. The (not canonical and gauge dependent) momentum $\hat p$ and Hamiltonian densities are
\bea
\hat{p}&=& -{\hat{m} (\dot{\hat{r}}+\ms N_r)\sqrt{\hat{g}_{rr}}}/ \sqrt{\ms N_t^2-(\dot{\hat{r}}+\ms N_r)^2}\,,\\
{\cal H}_{t}^\text{Shell}&=&\sqrt{\hat{p}^2 +\hat{m}^2 \hat{g}_{rr}} \delta(r-\hat{r})\,,~~~{\cal H}_{r}^\text{Shell}=-\hat{p} \delta(r-\hat{r})\,.\nonumber  
\eea
Combining the actions $S_\text{G}$ and $S_\text{Shell}$ we obtain the full Hamiltonian constraints ${\cal H}_{t,r}^\text{G}+{\cal H}_{t,r}^\text{Shell}=0$. These constraints simplify considerably in the gauge $\ms N_r=-\dot{\hat{r}}$, where $\hat{p}=S_{\text{Shell}}=0$. For a time-like  trajectory this corresponds to an observer  at rest with respect to the shell. With this choice we have the constraints
\be
0=A \pi_{\ms M}\pi_{\ms R}+A^{-1}{\ms M'\ms R'}+\hat{m} \delta(r-\hat{r})=\pi_{\ms R} \ms R'+\pi_{\ms M} \ms M'\,.
\ee
Integrating over the shell we find the discontinuity of the extrinsic curvature $\ms R'$ \cite{Israel:1966rt,bgg,fmp2}
\be\label{junction}
\hat{\ms R}'_+-\hat{\ms R}'_-=-\hat{m} G/ \hat{\ms R}\,,
\ee
where $\hat{\ms R}'_\pm \equiv \lim_{\epsilon\rightarrow 0}\ms R'(t,\hat{r}\pm \epsilon)$. The fixed discontinuity of the extrinsic curvature is inconsistent with its free variation within $\cal M$ that we assumed in the derivation of the gravitational action $S_\text{G}$. We therefore have to subtract this non-vanishing variation from the overall action in order to render the variational problem  well posed. This gives the boundary action $S_\text{BT}$ \cite{fmp2, Kraus:1994by}
\be
-\delta S_\text{BT}=\lim_{\epsilon\rightarrow 0}\left({\partial S_{\text{G}}\over \partial \ms R^\prime}\Big|_{r=\hat{r}-\epsilon} - {\partial S_{\text{G}}\over \partial \ms R^\prime}\Big|_{r=\hat{r}+\epsilon}\right) \delta \hat{\ms R}^\prime\,. 
\ee
To evaluate the boundary terms we require the gauge invariant gravitational action (\ref{classicalg}) for general curvatures, which we easily find by integrating the momentum (\ref{momM}),
\be
S_\text{G}\hspace{-.5pt}=\hspace{-1pt}\int_{r_\text{min}}^{r_\text{max}} \hspace{-2pt}\td r\hspace{-1pt}\Bigg[{{\ms R'}\cosh^{-1} \left({\ms R'\over \sqrt{\ms R'^2-A} } \right)-\sqrt{\ms R'^2-A}\over \eta_\pi G / \ms R}\Bigg]_{\hspace{-1pt}\ms R(t_\text{i},r)}^{\hspace{-1pt}\ms R(t_\text{f},r)} ,
\ee
where $\ms R(t_\text{i,f},r)$ are given by the boundary conditions. The boundary terms then become
\be\label{boundaryterm}
S_\text{BT}\hspace{-1pt}=\hspace{-1pt}\int_{\hat{\ms R}(t_\text{i})}^{\hat{\ms R}(t_\text{f})}\hspace{-1pt} {{\cosh^{-1}{\hat{\ms R}_+'\over \sqrt{\hat{\ms R}_+'^2-\hat{A}_+}}-\cosh^{-1}{\hat{\ms R}_-'\over \sqrt{\hat{\ms R}_-'^2-\hat{A}_-}}\over {\eta_\pi  G/\hat{\ms R}}}}\td \hat{\ms R}\,,
\ee
and we define the canonical momentum of the shell $S_\text{Shell}+S_\text{BT}\equiv \int \hat{P}\td \hat{R}$. Using  coordinates $R'_+=1$ we can solve the constraints for  the extrinsic curvature $R'_-$  in  (\ref{boundaryterm}) \footnote{Given the more convenient choice of variables for the gauge $g_{rr}=1$, the curvature $R'_-$ is most easily found using Eqs. (3.7), (3.13) and (3.14) of \cite{runaways}.}. Hamilton's equation for the shell reads
\be\label{ham}
\left({\td \hat{R}\over \td \ms T_+ }\right)^{-1}= {\td \hat{P}\over \td \ms M_+ }={\eta_\pi\over A_+}+\mathcal{O}(A_+^0)\,,
\ee
where in the last equality we expanded around the horizon $A_+\ll 1$. The last equality is what we naively expect for an in-falling particle: the velocity is negative (so $\eta_\pi=-1$) and approaches zero near the horizon. Combining (\ref{boundaryterm}) and (\ref{ham}) we have (just as above) the imaginary part of the action
\be
\text{Im}~S_\text{BT}\hspace{-0.3pt}\approx\hspace{-.3pt}\text{Im}~\int_{\ms M_-}^{\ms M_+}\td\ms M\int_{\hat{\ms R}(t_i)}^{\hat{\ms R}(t_f)}\td \hat {R} { \eta_\pi \hat{R} \over \hat{R}-{2 G \ms M} }= -\eta_\pi{\Delta \mathbb A_\text{h}\over 16G}.
\ee
Adding up all imaginary contributions we finally have 
\be\label{imaction}
2\times \text{Im}~S= {\Delta \mathbb A_\text{h}\over 4G}\,.
\ee
Note that the action is completely independent of the detailed properties of the shell, such as its tension $\sigma$ or whether it evolves along a light- or time-like trajectory. Therefore, it is not even relevant whether the in-falling matter is localized in a thin shell, or consists of a continuous collapsing cloud. The gauge invariant action associated to a change of the horizon radius of $\Delta \mathbb A_\text{h}$ is always given by (\ref{imaction}).

\section{Black Hole Entropy}
Having obtained the action for black hole formation in general relativity, we are now in a position to comment on three standard arguments to semiclassically associate black holes with an entropy. We will see that, in contrast to previous approaches, the gauge invariant action for general relativity requires no additional assumptions or prescriptions to compute the black hole entropy. We will not comment on the merit of the arguments or the interpretation and origin of the black hole entropy.

First, we can derive the black hole temperature by demanding the absence of a conical singularity in near-horizon geometry. A quantum field theory at finite inverse temperature $\beta$ is periodic in imaginary time, ${\ms  T}\sim  {\ms  T}+i{ \beta}$. Indeed, expanding the Schwarzschild metric  around the horizon, where a local observer experiences approximate Rindler space, we find a conical singularity unless the imaginary time  is periodically identified as ${\ms  T}\sim  {\ms  T}+4\pi i \ms R_\text{h}$. Assuming that (near the horizon) the mass $\ms M$ of the system generates Schwarzschild time $\ms T$ translations, we find the entropy ${\cal S}$ from the first law of thermodynamics
\be\label{bhentropy}
{\td \cal S\over \td \ms M}=\beta=4\pi \ms R_\text{h}\,,~~\text{or}~~\Delta S={\Delta  \mathbb A_\text{h}\over 4G}\,.
\ee

Second, we can derive the entropy by integrating the micro-canonical gravitational action (\ref{classicalg}) over a periodic Euclidean time $T\sim T+i \beta$  \cite{Brown:1992br,Brown:1992bq}
\be
S=- \int \td \ms M\int_0^{i\beta} \td \ms T  = -i \int \td \ms M \,8\pi G \ms M= -i {\mathbb A\over 4 G}\,,
\ee
where we are intentionally vague about the limits of integration. The entropy then is ${\cal S}=i S$.

Third, we can derive an entropy by relating the imaginary part of the action for emission or absorption of a spherical shell of mass $\Delta \ms M$ (as calculated in this work) to the Boltzmann factor \cite{KeskiVakkuri:1996xp,Parikh:1999mf}. The action of the in-falling matter is purely real, until the matter tunnels through the horizon of the outer spacetime. At that point, the action receives the imaginary contribution  (\ref{imaction}). The transmission probability through this classically forbidden region is  easily evaluated in the WKB approximation as
\be
{{\cal P}}_{I}=|\exp \left(i S\right)|^2=\exp \left(-{\Delta \mathbb A_\text{h}\over 4G}\right)\equiv e^{-\beta \Delta \ms M}\,.
\ee
Expanding the change in the area to leading order in the absorbed or emitted energy, $\Delta \mathbb A_\text{h}\approx 32\pi G^2 \ms M\Delta \ms M$ gives the inverse temperature $\beta=4\pi \ms R_\text{h}$, and the entropy just as in (\ref{bhentropy}).

Each of these three elegant derivations of the black hole entropy crucially required that the invariant energy ${\ms M}$ is the Hamiltonian associated to Schwarzschild time ${\ms  T}$ translations, which follows from the novel, coordinate independent action principle. In contrast, in other, coordinate dependent action principles the Hamiltonian is defined to coincide with the energy at the boundary, i.e. it is an assumption or prescription, not a result. This observation underlines the conceptual differences of a gauge invariant action principle.

\section{Discussion}
In this work we used the action (\ref{domwall1}) for thin shells in general relativity to show that physical black holes which form and decay have an entropy given by the Bekenstein-Hawking formula. Although the result finding was anticipated and may seem innocuous, if realized in nature, our coordinate independent approach has some profound conceptual implications that we now discuss.

 The entropy as defined here may be meaningful only for physical black holes: it parametrizes the ignorance about the information stored inside it during the formation process. The change in the horizon area and the entropy vanish for static black holes, as noted in \cite{tbgravity}. We can see directly from  (\ref{domwall1}) that there is meaningful entropy associated to the pure and static gravitational field: in the absence of  in-falling matter, $\hat{P}=0$,  the invariant mass is stationary, $\dot{\ms M}=0$, so the action vanishes. An eternal Schwarzschild black hole in flat space is not a thermal system, it -- by definition -- does not radiate or decay and instead is one unique and static state, consistent with the no-hair theorem.

The present action principle and the resulting entropy is fundamentally different from some other kinds of entropies that have  been associated with black holes. The reason is simple: commonly the  three-geometry at a boundary $\partial{\cal M}$ is fixed, while the present work instead is concerned with coordinate invariant boundary conditions. We can see the difference by comparing to the action
\be\label{ehaction}
S=S_\text{EH}+S_\text{GHY}\,,
\ee
which consists of the Einstein--Hilbert action and Gibbons--Hawking--York boundary terms, as discussed in \cite{Arnowitt:1962hi,York:1972sj,Regge:1974zd,Gibbons:1976ue,Brown:1992br,Brown:1992bq}. The variational principle $\delta S=0$ yields Einstein's equations only for fixed boundary three-geometries. This means that the lapse $N_t$ can be fixed at the boundary, $\delta N_t|_{\partial {\cal M}}=0$. Clearly, these boundary conditions are not coordinate independent, and not all charges associated with diffeomorphisms vanish, so for some observables (like vacuum decay rates in de Sitter space) additional input is required as to what coordinate choice is realized in nature at the boundary $\partial {\cal M}$. The action depends on this coordinate choice. A particularly popular prescription is to choose boundary terms such that the Hamiltonian (the charge associated to $t$-coordinate transformations) coincides with a parameter $M$, that is interpreted as the energy we desire the system to have. For a coordinate choice where $t$ agrees with the Schwarzschild time, $t\equiv \ms T$, the Hamiltonian  then -- by design -- is the  mass, and the action contains a term $S\supset -\int_{\partial {\cal M}}  M \td \ms T $. There are two issues with this. First, we could have chosen the Hamiltonian at the boundary to be anything, not just $M$, so this does not show but assume that $M$ is the  charge associated to Schwarzschild time translations. Second, even accepting the boundary term prescription as an additional law of nature,  $M$ would be the charge associated to time translations only at the boundary, instead of close to the horizon of a black hole as required in each of the semiclassical arguments for a black hole entropy. We conclude that in the absence of a fixed background metric the action (\ref{ehaction}) does not yield a well-defined time-translation structure. Wald emphasizes this largely ignored problem for de Sitter space in \cite{Wald:1999xu}. In contrast, the coordinate independent action we employ in this work has a well-defined time-translation structure everywhere, and thus yields unambiguous energies and entropies. All charges with respect to coordinate transformations vanish (including the total Hamiltonian), while the charge associated to Schwarzschild time translations is shown to be the invariant mass.

We briefly comment on the semiclassical computations in the related works \cite{KeskiVakkuri:1996xp,Parikh:1999mf}, in which the authors  argue that the action of an in-falling null shell, $S=\int \hat{P}\td \hat{R}$, alone gives rise to the tunneling probability (\ref{tunnelp}). We see from (\ref{domwall1}) that this would only give half of the desired action if Hamilton's equation $\td {\ms  M}/\td \hat{P}=\td \hat{\ms R}/\td {\hat{\ms T}}$ holds, i.e. if $\ms M$ is the Hamiltonian associated to Schwarzschild time. Instead, the authors of \cite{KeskiVakkuri:1996xp,Parikh:1999mf} modify Hamilton's equation to read $\td {\ms  M}/\td \hat{P}=\td \hat{\ms R}/\td {\hat{t}}$, where $\hat{t}$ is not the Schwarzschild time, but a particular choice for the time-coordinate that yields the known black hole entropy (\ref{tunnelp}). This shows how in a gauge-dependent approach one finds any desired result for the energy and entropy by a suitable choice of time-coordinate (akin to a measure problem).

To conclude, our work demonstrates how the conceptual novelties implied by demanding full diffeomorphism invariance can resolve some confusing aspects of semi-classical, isotropic gravity.

\section{Acknowledgments}
We thank Frederik Denef, Raphael Flauger, Daniel Green, Thomas Hartman, Matthew Kleban, Per Kraus, Juan Maldacena, Liam McAllister, John McGreevy, Ruben Monten and John Stout for useful discussions. This work was supported in part by DOE under grants no. DE-SC0009919 and by the Simons Foundation SFARI 560536.

\bibliographystyle{klebphys2}
\bibliography{bubblerefs}

\end{document}